\documentstyle[aps,prl]{revtex}
\begin{document}
\twocolumn

%
%
%

\narrowtext

\noindent {\bf Comment on ``Moving Glass Phase of Driven Lattices''}

In a recent Letter \cite{gld} Giamarchi and Le Doussal (GL) showed
that when a periodic lattice is rapidly driven through a quenched
random potential, the effect of disorder persists on large length
scales, resulting in a Moving Bragg Glass (MBG) phase. The MBG was
characterized by a finite transverse critical current and an array 
of static elastic channels.

They use a continuum displacement field ${\bf u}({\bf r},t)$, whose
motion (neglecting thermal fluctuations) in the laboratory frame obeys
$\eta\partial_tu_\alpha+\eta{\bf v}\cdot\nabla u_\alpha=c_{11}
\partial_\alpha\nabla\cdot{\bf u}+c_{66}\nabla^2u_\alpha+F_\alpha^p+F_\alpha
-\eta v_\alpha$, where $F_\alpha$ is the external driving force. As in
\cite{gld}, we choose $F_\alpha = F\delta_{\alpha,x}$ and denote by
$y$ the $d-1$ transverse directions. GL observe that the pinning force
$F^p_\alpha$ splits into {\it static} and {\it dynamic} parts,
$F^p_\alpha=F^{stat}_\alpha+F^{dyn}_\alpha$, with
$F^{stat}_\alpha({\bf r},{\bf u})=\rho_0V(r)\sum_{{\bf K}\cdot{\bf
v}=0} iK_\alpha e^{i{\bf K}\cdot({\bf r}-{\bf
u})}-\rho_0\nabla_\alpha V(r)$ and $F^{dyn}_\alpha({\bf r},{\bf u},t)=\rho_0V(r)\sum_{{\bf K}\cdot{\bf v}\not=0}
iK_\alpha e^{i{\bf K}\cdot({\bf r}-{\bf v}t-{\bf u})}$.
GL argue that in the sliding state at sufficiently large velocity
${\bf F}^{stat}$ gives the most important contribution to the
roughness of the phonon field $\bf u$, with only small corrections
coming from ${\bf F}^{dyn}$.  Since ${\bf F}^{stat}$ is along $y$ and
only depends on $u_y$, they assume $u_x=0$ and obtain a decoupled
equation for the transverse displacement $u_y$.  Analysis of this
equation then predicts the moving glass phase with the aforementioned
properties. 

In this Comment, we show that the model of Ref.~\onlinecite{gld}\
neglects important fluctuations that can destroy the periodicity in
the direction of motion. Following recent work by Chen et
al. \cite{chen}\ for driven charge density waves, it can be shown
\cite{mr} that the longitudinal {\it dynamic} force
$F^{dyn}_x$ does {\it not} average to zero in a coarse-grained model,
but generates an effective random static drag force $f_d({\bf
r})$. This arises physically from spatial variations in the impurity
density, and can be obtained by using a variant of the high-velocity
expansion or by coarse-graining methods. To leading order in ${1\over
F}$ its correlations are $<f_d({\bf r})f_d({\bf
0})>=\Delta_d\delta({\bf r})$, where $\Delta_d\sim \Delta^2/F$, and
$\Delta$ is the variance of the quenched random potential $V({\bf
r})$. The crucial difference from Ref.~\onlinecite{gld}\ is that
in contrast to ${\bf F}^{dyn}$, the effective static drag force
$f_d({\bf r})$ is strictly $\bf u$-independent, as guaranteed by the
precise time-translational invariance of the system coarse-grained on
the time scale $\sim 1/v$.

In the presence of $f_d$, we now reexamine both the elasticity and
the relevance of longitudinal dislocations (i.e. those with Burger's
vectors along $x$).  An improved elastic description begins with the
equation 
\begin{eqnarray}
\label{ourmodel}
\eta\partial_t u_\alpha+\eta{\bf v}\cdot\nabla u_\alpha &=&c_{11}
\partial_\alpha\nabla\cdot{\bf u}+c_{66}\nabla^2u_\alpha\nonumber \\
&+&\delta_{\alpha y}F_y^{stat}(u_y) + \delta_{\alpha x} f_d(\bf r)\;
\end{eqnarray}
Surprisingly, a simple calculation leads to a transverse correlator,
$B_y({\bf r})=<[u_y({\bf r})-u_y({\bf 0})]^2>$, that is (for $d>1$)
asymptotically identical to that found by GL, which exhibits highly
anisotropic logarithmic scaling for $d\leq3$.  In contrast, the $u_x$
roughness is dominated by $f_d$, and $B_x({\bf r})=<[u_x({\bf
r})-u_x({\bf 0})]^2>$  grows algebraically,
$\sim(\Delta_d/c_{66}^2) r^{4-d}$ for $d<4$ and $x<c_{66}/\eta v$,
crossing over for $x>c_{66}/\eta v$ (and $d<3$) to
$B_x({\bf r})\sim (\Delta_d/c_{66}\eta v)y^{3-d} H(c_{66} x/\eta v
y^2)$, with $H(0)=\mbox{const.}$ and $H(z>>1)\sim z^{(3-d)/2}$.  We
stress that because of $\bf u$-independence of $f_d$ this power-law
scaling for $B_x(\bf r)$ holds out to arbitrary length scales, in
contrast to that for $B_y(\bf r)$ valid only in the Larkin regime as
lucidly discussed by GL.\cite{gld} Thus, even within the elastic
description, translational correlations along $x$ are short-ranged
(stretched exponential). Stability with respect to dislocations is
more delicate.  Nevertheless, arguments analogous to those of
Ref.\onlinecite{BF}\ suggest that dislocation unbinding will occur for
$d\leq 3$, converting the longitudinal spatial correlations to the
pure exponential (liquid-like) form.  We stress that this situation
corresponds not to $u_x=0$, as assumed in Ref.\onlinecite{gld}, but
rather $\langle u_x^2 \rangle = \infty$ (indeed, $u_x$ is {\sl
multivalued}).

We therefore argue that for intermediate velocities (for $d\leq 3$) a
moving vortex solid is organized into a stack of {\em liquid}
channels, i.e. it is a moving {\em smectic}. This is in agreement with
structure functions and real-space images from recent
simulations\cite{moon}.  The model for this nonequilibrium smectic
state will be the subject of a future publication\cite{mr}.  An
interesting possibility is that at {\sl very} large velocities,
nonequilibrium KPZ type nonlinearities(as in Ref.\onlinecite{chen})
might lead to a further transition to a more longitudinally ordered
state, with rather different underyling physics from the MBG.

MCM, LR, and LB were supported by the National
Science Foundation, grants No. DMR-9419257, No. DMR-9625111,
and No. PHY94--07194, respectively.

{\small \noindent Leon Balents$^1$, M. Cristina Marchetti$^2$,  and Leo
Radzihovsky$^3$ \\
\indent$^1$Institute for Theoretical Physics \\
\indent\hspace*{1pt}University of California \\
\indent\hspace*{1pt}Santa Barbara, CA 93106--4030\\
\indent$^2$Physics Department\\
\indent\hspace*{1pt}Syracuse University\\
\indent\hspace*{1pt}Syracuse, NY 13244\\
\indent$^3$Physics Department\\
\indent\hspace*{1pt}University of Colorado\\
\indent\hspace*{1pt}Boulder, CO 80309\\
}


\end{document}